\documentclass[a4paper,UKenglish,cleveref, autoref, thm-restate]{lipics-v2021}

\usepackage{braket}
\usepackage{tikz}
\usetikzlibrary{matrix}

\title{Set membership with two classical and quantum bit probes} 

\titlerunning{Set membership with two probes} 

\author{Shyam S. Dhamapurkar}{Southern University of Science and Technology, Shenzhen, P.R.~China}{shyam18596@gmail.com}{}{}

\author{Shubham Vivek Pawar}{Hubli, Karnataka, India}{shubham.pawar@students.iiserpune.ac.in}{}{Partial work was done while interning at TCS Innovation Lab, Mumbai}

\author{Jaikumar Radhakrishnan}{Tata Institute of Fundamental Research, Mumbai, India}{jaikumar@tifr.res.in}{}{Supported by the Department of Atomic Energy, Government of India, under project no.~RTI4001.}

\authorrunning{S.~Dhamapurkar, S.~Pawar, J.~Radhakrishnan} 

\Copyright{Shyam Dhamapurkar, Shubham Pawar, Jaikumar Radhakrishnan} 



\category{ } 

\relatedversion{} 

\usepackage{tikz}
\usepackage[linesnumbered,algoruled,boxed,lined]{algorithm2e}
\nolinenumbers
\begin{document}

\maketitle

\begin{abstract}
We study the classical and quantum bit-probe versions of the static set membership problem :  Given a subset, $S$ $(|S| \leq n)$ of a universe, $\mathcal{U}$ $(|\mathcal{U}| = m\gg n)$, represent it as a binary string in memory so that the query ``Is x in S?'' ($x\in \mathcal{U}$) can be answered by making at most $t$ probes into the string. Let $s_{A}(m,n,t)$ denote the minimum length of the bit string 
in any scheme that solves this static set membership problem. We show that for $n\geq 4$
\[ s_A(m,n,t=2) =
\begin{cases}
    \mathcal{O}(m^{1-1/(n-1)}) & \text{if } n = 0 \pmod{3};\\
    \mathcal{O}(m^{1-1/n})     & \text{if } n = 1,2 \pmod{3};\\
    \mathcal{O}(m^{6/7})       & \text{if } n = 6,7.
\end{cases}
\]
These bounds are shown using a common scheme that is based on a graph-theoretic observation on orienting the edges of a graph of high girth. For all $n\geq 4$, these bounds substantially improve on the previous best bounds known for this problem, some of which required elaborate constructions~\cite{DBLP:journals/tcs/BaigK20a}. Our schemes are explicit. A lower bound of the form $s_A(m,n,2)=\Omega(m^{1-\frac{1}{\lfloor{n/4}\rfloor}})$ is known for this problem; better lower bounds are known for specific small values of $n$~\cite{BAIG2020543, DBLP:conf/iwoca/BaigKS19, DBLP:conf/walcom/BaigK18,DBLP:conf/walcom/BaigKS19,DBLP:journals/tcs/BaigK20a}. 

We consider the quantum version of the problem, where access to the bit-string $b \in \{0,1\}^s$ is provided in the form of a quantum oracle that perform the transformation $\mathcal{O}_b: \ket{i} \mapsto (-1)^{b_i} \ket{i}$. Let $s_Q(m,n,2)$ denote the minimum 
length of length of the bit string that solves the above set membership problem in the quantum model (with adaptive queries but no error). 
We show that for all $n\leq m^{1/8}$, we have $s_{QA}(m,n,2) = \mathcal{O}(m^{7/8})$.
This upper bound makes crucial use of Nash-William's theorem~\cite{diestel2000} for decomposing a graph into forests. This result is significant for because, prior to this work, it was not known if quantum schemes yield any advantage over classical schemes. We also consider schemes that make a small number of quantum \emph{non-adaptive} probes. In particular, we show that the space required in this case, $s_{QN}(m,n=2,t=2)=O(\sqrt{m})$ and $s_{QN}(m,n=2,t=3) = O(m^{1/3})$; in contrast, it is known that two non-adaptive classical probes yield no savings. Our quantum schemes are simple and use only the fact that the XOR of two bits of memory can be computed using just one quantum query to the oracle.
\end{abstract}

\newpage

\section{Introduction}
\label{sec:typesetting-summary}
We consider the problem of representing small subsets $S$ of a universe $[m]=\{1,2,\ldots,m\}$ in memory as a bit string so that membership queries of the form ``Is $x \in S$?'' can be answered with a small number of bit probes to the memory~\cite{BMRV}. One natural way of representing sets in memory is the characteristic vector, which uses $m$ bits of memory and answers membership queries using a single bit probe. Since there are only $O({m \choose n})$ sets of size at most $n$ (assume $n\ll m$) one might hope to represent them using $O(\log_2 {m \choose n})\approx O(n\log m)$ bits of memory.  However, compression to near the information-theoretic limits comes with a cost: membership can no longer be determined by reading just a small number of bits. To describe the trade-offs between efficiency of compression and the effort for extraction (measured as the number of bit probes), we will use the following notation~\cite{MR17}.  Let $s_{N}(m,n,t)$ denote the minimum number of bits in a scheme that can represent sets of size at most $n$ and answer membership queries by probing at most $t$ bits of the memory non-adaptively (that is, the probes are made in parallel). We write $s_{A}(m,n,t)$ if the scheme is adaptive; we use the subscript $Q$ if the scheme makes quantum queries (zero-error), which can be adaptive or non-adaptive~\cite{DBLP:conf/focs/RadhakrishnanSV}, and write $s_{QA}(m,n,t)$ and $s_{QN}(m,n,t)$. Clearly,
\[ s_{N}(m,n,t) \geq \left\{\begin{array}{l} 
s_{A}(m,n,t)\\
s_{QN}(m,n,t)
\end{array}\right\}
\geq s_{QA}(m,n,t)
\]
Radhakrishnan, Sen and Venkatesh~\cite{DBLP:conf/focs/RadhakrishnanSV} obtained lower bounds, which for the range of parameters of interest to us ($n\leq \sqrt{m}$, $t$ constant) implies the following. 
\[ s_{QA}(m,n,t) = \Omega(m^{1/t}n^{1-1/t}).\]
(A similar lower bound in the classical setting was shown by Buhrman \emph{et al.}~\cite{BMRV}.) Note that this bound shows that if the space is compressed to $O(n \log m)$, then $t=\Omega(\log m)$. Furthermore, if $t=1$ no compression is possible even for $n=1$; it also shows that $s_{Q}(m,1,2)=\Omega(\sqrt{m})$, for which there is a matching upper bound
$s_{N}(m,n=1,2)=O(\sqrt{m})$. The first, non-trivial case is $n=2$ and $t=2$, where the above bound implies that $s_{QA}(m,2,2) =\Omega(\sqrt{m})$. It is known that this bound is not tight for classical schemes; better lower bound are known: $s_{N}(m,2,2)=m$ and $s_{A}(m,2,2)=\Omega(m^{4/7})$~\cite{BMRV, RSS10}.
Remarkably, it is known that $s_{A}(m, 3, 2) = \Theta(m^{2/3})$~\cite{kesh:LIPIcs:2018:9911,BAIG2020543}.
Two-probe classical schemes have been constructed for representing small sets in several works 
starting with Alon and Feige~\cite{DBLP:conf/soda/AlonF09} 
(see, e.g.,~\cite{DBLP:journals/tcs/BaigK20a,DBLP:conf/walcom/BaigK18, DBLP:conf/walcom/BaigKS19,DBLP:journals/corr/0003R15a,kesh:LIPIcs:2018:9911,DBLP:conf/esa/LewensteinMNR14, DBLP:conf/esa/RadhakrishnanRR01}, where sets of specific sizes are considered); the following upper and lower bounds was obtained by Garg and Radhakrishnan~\cite{DBLP:journals/corr/0003R15a}.
\begin{equation} \label{eq:garg:radhakrishnan}
  \Omega(m^{1-\frac{1}{\lfloor{n/4}\rfloor}})  \leq s_{A}(m,n,t=2) \leq 
        \mathcal{O}(m^{1-\frac{1}{4n+1}}). 
\end{equation}
which roughly characterizes the space requirement for the problem, and, in particular, establishes that no savings over the standard characteristic vector representation can be expected if $n \geq \log m$. We show the following.
\begin{theorem}[Result 1] \label{thm:result1}
For $n\geq 4$,
\[ s_A(m,n,t=2) =
\begin{cases}
    \mathcal{O}(m^{1-1/(n-1)}) & \text{if } n = 0 \pmod{3};\\
    \mathcal{O}(m^{1-1/n})     & \text{if } n = 1,2 \pmod{3};\\
    \mathcal{O}(m^{6/7})       & \text{if } n = 8,9.
\end{cases}
\]
\end{theorem}
The above bounds improve the bounds for all other values of $n$ (see \cref{fig:comparison}) for a comparison. Unlike the previous works, where different constructions (some of which quite intricate) were invented for different set sizes, our result is obtained using a unified approach based on graphs of high girth. (For $n=2,3$, the construction matches the currently best bounds.) To obtain a scheme for a set given size, one just plugs in the best available result for high-girth graph and obtains the bound claimed above. More importantly for us, the method used here inspires a quantum scheme that yields the following bounds.
\begin{theorem}[Result 2] \label{thm:result2}
 $s_{QA}(m,n=m^{1/8},t=2) = \mathcal{O}(m^{7/8})$.
\end{theorem}
This result is significant because it shows that, unlike in the classical case, two probes give substantial savings over the characteristic vector representation for sets substantially larger than $\log m$ (see the remark above
following \cref{eq:garg:radhakrishnan}). Before this work, quantum schemes were not known to provide significant savings over classical schemes. Our quantum scheme is also based on dense graphs that are locally sparse, this time we do not make use of high girth. Instead, we 
invoke a result of Nash-Williams~\cite{diestel2000} on covering the edges of a graph with two forests. After this, our construction uses only the following basic fact from quantum computation (Deutsch's algorithm~\cite{DBLP:books/daglib/0046438}): the parity of two bits of memory can be computed using just one quantum probe. In fact, only the second probe in our scheme is truly quantum.
This result opens the possibility that the lower bounds
of $\sqrt{mn}$ cited above is perhaps achievable for quantum schemes. We show in fact that for $n=2$, the lower bound can be matched using \emph{non-adaptive} constructions.
\begin{theorem}[Result3] \label{result:quantumbounds}
\begin{align*}
    s_{QN}(m,n=2,t=2) &= \mathcal{O}(\sqrt{m}); \\
    s_{QN}(m,n=2,t=3) &= \mathcal{O}(m^{1/3}).
\end{align*}
\end{theorem}
The query scheme is simple. The query scheme for $(n=2,t=2)$ on input $x$ computes for locations $\ell_1(x), \ell_2(x), \ell_3(x), \ell_4(x)$, and returns `Yes' iff the bits at the first two locations are different and the bits at the last two locations are different, that is, we use an AND of two 
inequality computations, each of which requires just one quantum probe. A similar query scheme that uses an AND of three inequality computations gives the three-probe non-adaptive quantum scheme.
We also obtain non-adaptive two-probe schemes with sublinear space for storing sets with more elements (see~\cref{appendix:non-adaptive}). These bounds are interesting because no non-adaptive two-probe classical scheme exists with sublinear space~\cite{BMRV}.



      

\section{Classical two-probe adaptive schemes}
\label{sec:classical_adaptive}
In this section we establish~\cref{thm:result1}. Our two-probe adaptive schemes are based on dense graphs of high girth. We first specify the storage scheme and the query schemes based on an underlying graph. Then, to complete the proof, we will show the following: (i) if the underlying graph has high girth, then there is an assignment of values to the memory such that all queries are answered correctly; (ii) the available explicit
constructions of dense graphs of high girth in the literature yield the claimed bounds.

    

     

     
\begin{definition}[Classical $(G,K)$-scheme]
Let $G$ be a directed graph with $N$ vertices and $M$ edges; let $K$ be a positive integer.
We refer to the following as a $(G,K)$-scheme. The storage consists of two bit arrays, $A$ and $B$. To answer a membership query the decision tree will make the first probe to array $A$ and the second probe to array $B$.
\begin{description}
\item[Edge array:] An array $A: E(G) \rightarrow \{0,1\}$, indexed by edges of $G$. 
\item[Vertex array:] A two dimensional array $B: V \times [K] \rightarrow \{0,1\}$. 
\item[Elements:] We identify our universe of elements $[m]$ with
a subset of $E(G) \times [K]$ (so we must ensure that the graph has
at least $m/K$ edges); thus, each element $x \in [m]$ will be referred to as $(e,i)$. 
\item[Query:] We represent an edge of $G$ as an ordered pair of the form $e=(v_0,v_1)$ with the convention that $v_0 < v_1$. To process the query for the element $x=(e,i)$, we read $A[e]$ (first probe); then 
we return the value $B[v_{A[e]},i]$ (by making the second probe into $B$). In other words, we may think of  $A[e]$ as a bit that orients the edge $e$ towards either its smaller vertex or the larger vertex; depending on this bit, the second probe is made into the array $B$ corresponding to the vertex towards which the edge points. Note that this scheme is adaptive: the second probe depends on the first.
\item[Space:] We will ensure that $MK \geq m$. The space used by this scheme is then $NK + M$ bits ($NK$ for the $N$ vertex array and $K$ for the edge array). By choosing the graph $G$ and the parameter $K$ appropriately we will show that our schemes use small space.
\end{description}
\end{definition}

The following lemma provides the connection between dense graphs of high girth and efficient two-probe adaptive schemes.
\begin{lemma} \label{lm:classical_adaptive_main}
Let  $G$ be a graph with $N$ vertices $M$ edges and girth $g$ such that
$n\leq \lfloor\frac{3g}{4}\rfloor$ and $M \leq m$.
Then, $s_{A}(m,n,2) \leq M + N\lceil m/M \rceil$.
\end{lemma}
Before we present the proof of this lemma formally, let us derive from it the bounds claimed in \cref{thm:result1}. Since every graph has a bipartite subgraph that includes at least half the edges, it is sufficient to restrict attention to bipartite graphs, and hence to graphs whose girth is even. The smallest even number $g$ such that $n\leq \lfloor\frac{3g}{4}\rfloor$ is given by
\begin{equation}
    g(n) = \begin{cases}
        4 \lceil{n/3}\rceil & n = -1,0 \pmod{3};\\
        4 \lceil{n/3}\rceil - 2& n = 1 \pmod{3}.
    \end{cases}
\end{equation}
Now, suppose that for a girth $g$, there are constant $c(g)$ $d(g)$ and $\tau(g)$, such that for all large $L$, there is a graph with at most $c(g)L$ vertices, girth $g$ and $d(g) L^{1+\tau(g)}$ edges. Then, taking a graph with $N=\Theta(m^{1/(1+2\tau(g(n)))})$ vertices and $M=\Theta(m^{(1+\tau)/(1+2\tau)})$ edges
in~\cref{lm:classical_adaptive_main}, we obtain the following.
\begin{corollary} \label{cor:classical_adaptive}
$s_{A}(m,n,2) = \mathcal{O}(m^{(1+\tau(g(n)))/(1+2\tau(g(n)))})$.
\end{corollary}
In particular, by using the current best constructions of dense graphs with large girth we obtain the following bounds for $s_{A}(m,n,2)$. For example, we may take $\tau(6)=1/2$, $\tau(8)=1/3$ and $\tau(12)=1/5$ based on graphs described Wenger~\cite{WENGER1991113}.
(In \cref{sec:appendix_explicit}, we explain in what sense these constructions, and hence the resulting schemes, are explicit.)
\begin{figure}
\begin{center}
 \begin{tabular}{|c|c|c|c|c|l|}
 \hline
$n$ & girth $g(n)$ & $\tau(g(n))$ &Our bound ($m^{(1+\tau)/(1+2\tau)}$)& Previous best bound\\
\hline
 2,3 & 4 & 1 & $\mathcal{O}(m^{2/3})$ & $\mathcal{O}(m^{2/3})$ ~\cite{kesh:LIPIcs:2018:9911} \\
 4  & 6 & 1/2 & $\mathcal{O}(m^{3/4})$ (using~\cite{WENGER1991113}) & $\mathcal{O}(m^{5/6})$~\cite{DBLP:conf/iwoca/BaigKS19,DBLP:journals/tcs/BaigK20a}\\
 5,6& 8 & 1/3 & $\mathcal{O}(m^{4/5})$ (using~\cite{WENGER1991113})& $\mathcal{O}(m^{5/6})$~(for $n=5$ \cite{DBLP:journals/tcs/BaigK20a})\\
 7 & 10 & 1/5 & $\mathcal{O}(m^{6/7})$ (using~\cite{benson_1966})& $\downarrow$\\
 8,9 & 12 & 1/5 & $\mathcal{O}(m^{6/7})$ (using~\cite{benson_1966})&$\downarrow$ \\
 $n=3r-2 $& $4r-2$& $1/(3r-4)$ & $\mathcal{O}(m^{1-\frac{1}{n}})$ (using \cite{girthpaper})& $\mathcal{O}(m^{1-1/(4n+1})$~\cite{GR15}\\
 $n=3r-1,3r$& $4r$& $1/(3r-3)$&  $\mathcal{O}(m^{1-\frac{1}{n}})$, $\mathcal{O}(m^{1-\frac{1}{n-1}})$ (using \cite{girthpaper})& $\mathcal{O}(m^{1-1/(4n+1})$~\cite{GR15} \\
 \hline
\end{tabular}
\end{center}
\caption{Our upper bounds use explicit constructions of graphs of large girth available in the literature (see~\cref{sec:appendix_explicit})}
\label{fig:comparison}
\end{figure}

\begin{proof}[Proof of \cref{lm:classical_adaptive_main}]
Fix a graph $G$ with $N$ vertices and $M$ edges as in the statement of the theorem. Consider the $(G,K)$-scheme with $K=\lceil{m/M}\rceil$. Clearly the space used by the scheme is $N + KM = N + M \lceil{m/M}\rceil$.
It remains to show that there is an assignment to the edge and vertex arrays of this scheme so that every query is answered correctly. Fix a set $S$ of at most $n$ elements; recall that the elements of the universe have the form $(e,i)$, where $e$ is an edge of the graph and $i\in [K]$. We will assign values to the two arrays in two steps. First the edge array $A$ will be assigned values. Recall that this assignment corresponds to assigning directions to the edges. We will show below
how this is to be done. Assuming this we show how the array $B$ is initialized. To start with, we initialize array $B$ with zeros. Now for each element $(e,i) \in S$ (say $e=(v_0,v_1)$ where $v_0 < v_1$), if $A[e]=0$ we set $B[v_0,i]=1$, otherwise we set $B[v_1,i]=1$. This assignment ensures that the query scheme described above will answer correctly for each element in $S$, so there are no \emph{false negatives}, no matter what initial orientation of the edges is chosen. The key idea is to choose an orientation that avoids \emph{false positives}; we must ensure that the value in the array $A$ are set in such a way that an element not in $S$ and an element in $S$ 
do not make second probes to the same location in array $B$. 
\cref{defn:safe} below formally describes such a \emph{safe orientation}. (Here edges $e$ such that $(e,i) \in S$ for some $i$ are colored GREEN and the other edges are colored RED. Thus, there are at most $n$ GREEN edges.) Then, \cref{lm:safe} below shows that the graph $G$ above has a safe orientation. It follows that our query scheme answers all the queries correctly.
\end{proof}

\begin{definition}[Safe orientation]~\label{defn:safe}
Suppose $H$ is a graph whose edges are colored RED and GREEN. We say that an orientation of edges is \emph{safe} if every vertex with an incoming GREEN edge has only one edge coming into it. 
\end{definition}




\begin{lemma}\label{lm:safe}
    Suppose $H$ is a graph with even girth $g\geq 4$ and $n \leq \lfloor{3g/4}\rfloor$ GREEN edges.  Then, $G$ has a safe orientation.  (This claim should have a simple proof, but we have not been able to find one that covers all cases succinctly.)
\end{lemma}



\smallskip

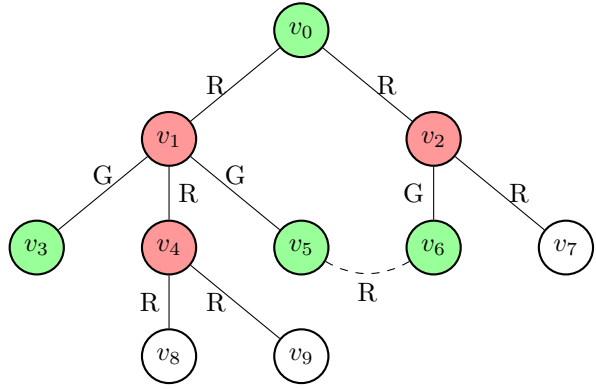
\begin{figure}
    \centering
    \begin{tikzpicture}
\matrix[matrix of math nodes, anchor= center,
      nodes={circle, draw, minimum size = 0.2 cm},
        column sep = {1 cm},
        row sep={0.7 cm},thick]
{ &  & |(a) [fill=green!40!white]| v_{0} & &  \\
  & |(b)[fill=red!40!white]|v_{1} &  & |(c)[fill=red!40!white]| v_{2}  \\
|(d)[fill=green!40!white]| v_{3} & |(e)[fill=red!40!white]|v_{4} & |(f)[fill=green!40!white]|v_{5} & |(g)[fill=green!40!white]|v_{6}& |(h)|v_{7} \\             
& |(i)| v_{8} & |(j)|v_{9}\\
};
  \foreach \from/\to/\weight/\where 
  in {a/b/R/left, a/c/R/right, b/d/G/above,
  b/e/R/right, b/f/G/above, c/g/G/left, c/h/R/right, e/i/R/left, e/j/R/left}
     \draw (\from) to [-]  node[midway,\where]  
     {{\weight}}  (\to);
    \draw[dashed] (f) to [out=-30,in=210] [-] node[midway,below]{R} (g);   
\end{tikzpicture}
    \caption{The edge is $(v_5,v_6)$ is a blocking edge
    and $(v_0,v_1,v_5,v_6,v_2,v_0)$ is the resulting 
    GREEN-dominated cycle, even though it has more RED edges than GREEN edges}
    \label{fig:green-dominated}
\end{figure}

\noindent \emph{Preliminaries:} In the following, suppose $H$ is a 
graph with even girth $g$ and $n \leq \lfloor{3g/4}\rfloor$ GREEN edges. To find the necessary orientation, we will proceed by induction on the size of $H$ (its total number of edges plus vertices). For the base case, note that a graph with no edges clearly has a safe orientation. For the induction step, we will identify an initial set of vertices $V'$ such that all edges that have at least one end point in $V'$ can be safely oriented towards a vertex in $V'$. We then delete $V'$ and the edges incident on it, and use induction to extend this orientation to the rest of $H$. To identify the set $V'$, we will use a \emph{breadth first search} procedure formally described below. This procedure produces a \emph{breadth first search tree} or a \emph{breadth first search forest} as usual, but we need to impose the following condition on it.
\begin{quote}
If a vertex $w$ in the tree is connected to its parent by a RED edge, then all of $w$'s children are connected to $w$ using GREEN edges; thus, in any root to leaf path in the tree, RED edges do not appear consecutively. 
\end{quote}
To enforce this, when a RED edge is added to the tree, we will mark
the vertex it leads to RED; then when we visit this vertex, we only explore vertices connected to it by GREEN edges. If a GREEN edge is added to the tree, we mark the new vertex GREEN; then when we visit this vertex, we explore all its edges, whether GREEN or RED. The
formal code is presented in \cref{alg:BFS}. (This is reminiscent of the breadth first search procedure employed by certain matching algorithms to discover augmenting paths; there one alternately explores either only the matched edge or all edges).
\begin{algorithm}
\caption{Breadth-First Search (BFS)}
\label{alg:BFS}
\SetKwInOut{KwIn}{Input}
\SetKwInOut{KwOut}{Output}
\KwIn{A non-empty subset $Z \subseteq V(H)$}

\KwOut{A BFS forest rooted at the vertices in $Z$}
$Q$ = empty queue \;
push all elements of $Z$ into $Q$ and mark them {GREEN}\;
\While{$Q$ is non-empty}{
    $v$ = pop($Q$)\;
    \If{$v$ is marked \rm{GREEN}}{
        push all unmarked neighbors $w$ of $v$ into $Q$\;
        \tcp{now assign them colors as follows}
        \If{$\{v,w\}$ is \rm{GREEN}}{add $\{v,w\}$ to the forest, and mark $w$ \rm{GREEN}}
        \Else{mark $w$ \rm{RED}}
        }
    \If{$v$ is marked \rm{RED}}{
       push all unmarked neighbors $w$ of $v$ with $\{v,w\}$ GREEN into $Q$\;
        add $\{v,w\}$ to the forest and mark $w$ GREEN\;
        }
    
    }

\end{algorithm}
As a first attempt we might want to orient the edges of the forest away from the roots and hope to extend this orientation to the other edges that have at least one end point in the forest. If this gives a valid orientation for these edges, we let $V'$ be the vertex set of the forest, and proceed as above. Unfortunately, this straightforward method may run into problems; this motivates the following definition.
\begin{definition}[Blocking edge, see \cref{fig:BFStree}]
In the forest constructed by BFS, we say that a non-tree edge is a blocking edge if (i) it is a non-tree {\rm{GREEN}} edge both of whose end points are visited during BFS, or (ii) it is a non-tree {\rm{RED}} edge with both end points marked {\rm{GREEN}}.
\end{definition}
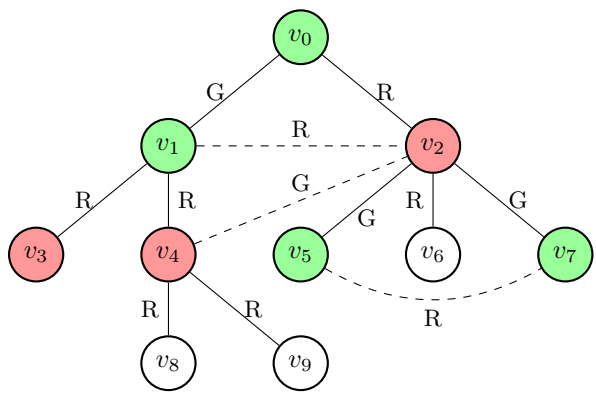
\begin{figure}[h]
    \centering
    \begin{tikzpicture}
\matrix[matrix of math nodes, anchor= center,
      nodes={circle, draw, minimum size = 0.2 cm},
        column sep = {1 cm},
        row sep={0.7 cm},thick]
{ &  & |(a) [fill=green!40!white]| v_{0} & &  \\
  & |(b)[fill=green!40!white]|v_{1} &  & |(c)[fill=red!40!white]| v_{2}  \\
|(d)[fill=red!40!white]| v_{3} & |(e)[fill=red!40!white]|v_{4} & |(f)[fill=green!40!white]|v_{5} & |(g)|v_{6}& |(h)[fill=green!40!white]|v_{7} \\             
& |(i)| v_{8} & |(j)|v_{9}\\
};
  \foreach \from/\to/\weight/\where 
  in {a/b/G/left, a/c/R/right, b/d/R/left,
  b/e/R/right, c/f/G/below, c/g/R/left, c/h/G/right, e/i/R/left, e/j/R/right}
     \draw (\from) to [-]  node[midway,\where]  
     {\small{\weight}}  (\to);
    \draw[dashed] (f) to [out=-30,in=210] [-] node[midway,below]{\small{R}} (h);   
    \draw[dashed] (b) to [-] node[midway,above]{\small{R}} (c);  
    \draw[dashed] (e) to  [-] node[midway,above]{\small{G}} (c);   
\end{tikzpicture}
    \caption{The BFS tree: $(v_1,v_2)$ is not a blocking edge, $(v_2,v_4)$ and $(v_5,v_7)$ are blocking edges}
    \label{fig:BFStree}
\end{figure}
We will see that if there are no blocking edges, then the above strategy will work; otherwise, $H$ has a cycle with many GREEN edges, and we will be able to exploit that.
\begin{definition}[GREEN-diminated cycle, see \cref{fig:green-dominated}]
We say that a cycle in $H$ is \emph{\rm{GREEN}-dominated} if all but (perhaps) one of its RED edges are followed by a \rm{GREEN} edge.
\end{definition}
We will establish the following two lemmas below.
\begin{lemma}\label{lm:hasnocycle}
Suppose $H$ has no \rm{GREEN}-dominated cycle.
Suppose $V'$ is the set of vertices of $H$ visited by BFS starting at a vertex $v_0$. Then there is a safe orientation of edges of $H$ incident on $V'$.
\end{lemma}

\begin{lemma}\label{lm:hascycle}
Suppose $H$ has an \rm{GREEN}-dominated cycle $C$. Let $H'$ be the graph obtained by deleting from $H$ all edges of $C$. Let $V'$ be the vertices visited by BFS in $H'$ starting with the vetex set $V(C)$ of the cycle $C$. Then, there is a safe orientation of edges of $H$ incident on $V'$. 
\end{lemma}
Let us use these lemmas to complete the proof of~\cref{lm:safe}.
\begin{proof}[Proof of the \cref{lm:safe}]
If $H$ has no GREEN-dominated cycle, then by \cref{lm:hasnocycle}, we obtain an initial set of vertices $V'$ and an orientation of all edges incident on it.
If $H$ has a GREEN-dominated cycle, then by \cref{lm:hascycle} we obtain an initial set of vertices $V'$ and an orientation of all edges incident on it.
We extend this orientation to the remaining edges of the graph by deleting $V'$ and all edges incident on it, and applying induction to the resulting subgraph
induced by the vertex set $V\setminus V'$.
\end{proof}
We now return to the unproved lemmas.
\begin{proof}[Proof of \cref{lm:hasnocycle}]
Let $v_0$ be an arbitrary vertex. Consider the tree produced by BFS starting with $Z=\{v_0\}$. We claim that there is no blocking edge for the resulting tree. For suppose $e=\{a,b\}$ is a blocking edge. Let $v$ be the least common ancestor of $a$ and $b$, and recall that in the paths that connect $v$ to $a$ and $v$ to $b$, no RED edge is followed by a RED edge. Let $C$ be the cycle formed by taking the path from $v$ to $a$ followed by $e$ and then the path from $b$ back to $v$. If $e$ is GREEN, then this cycle is GREEN-dominated, contradicting our assumption. If $e$ is RED, then by the definition of blocking edge, both $a$ and $b$ are marked GREEN, that is, the tree edges connecting them to their parents are GREEN (note that $e$ is not a back edge because both its vertices are GREEN). Again the cycle $C$ is GREEN-dominated, contradicting our assumption. Thus, the tree has no 
blocking edges. Let $V'$ be the vertices visited by BFS. Orient all tree edges away from the root $v_0$. The remaining edges incident on $V'$ (which cannot be GREEN)  have at least one vertex marked RED, because they are not blocking. Orient them towards that RED vertex. It can be verified that the GREEN edges that received an orientation are all tree edges, and are oriented towards distinct GREEN vertices. The RED edges are oriented towards RED vertices. So all edges incident on $V'$ can be oriented safely.
\end{proof}

\begin{proof}[Proof of \cref{lm:hascycle}]
Fix a \rm{GREEN}-dominated cycle $C$ in $H$. Suppose it has $\ell_1$ edges (for some $\ell_1 \geq g$) of which say $n_1$ are GREEN. Then,
\begin{equation}
    n_1 \geq \lceil{(\ell_1-1)/2}\rceil \geq g/2, \label{eq:cycle1} 
\end{equation}
because $g$ is even. First, suppose the resulting BFS forest has no blocking edges, then let $V'$ be the vertices of this BFS forest. We orient the edges in $C$ so that it becomes a directed cycle (we may choose either of the two ways to do this). Then, we orient all tree edges away from the roots in the BFS forest. Note that all other edges incident on $V'$ must necessarily be RED; each such edge has at least one RED vertex in $V'$. We orient each such edge towards a RED vertex, and obtain the desired safe orientation. 

Next, suppose thee is a blocking edge $e=\{a,b\}$. If $a$ and $b$ belong to the same tree of the forest, then $e$ and the paths from $a$ and $b$ to their least common ancestor form a \rm{GREEN}-dominated cycle, consisting of say $\ell_2 \geq g$ edges of which $n_2$ are GREEN. Then,
\begin{equation}
    n_2 \geq  \lceil (\ell_2-1)/2 \rceil\geq g/2. \label{eq:cycle2} 
\end{equation}
From \cref{eq:cycle1} and \cref{eq:cycle2}, we obtain, $n \geq n_1 + n_2 \geq g$, contradicting our assumption that $n \leq 3g/4$. So, we may assume that $a$ and $b$ belong to different trees of the forest. Then, travelling from the root $r_1$ of $a$'s tree to $a$, crossing over along $e$ to $b$, and then travelling to the root $r_2$ of the tree of $b$, we obtain a path $P^*$, where no RED edge is followed by a RED edge; in particular, every RED edge except perhaps the last, is followed by a RED edge. Suppose this path has $\ell_3$ edges of which $n_3$ are GREEN. We have the following.
\begin{align}
    \ell_3 &\geq g - \lfloor{\ell_1/2}\rfloor; && \mbox{($G$ has girth $g$)}\label{eq:cycle3}\\
    2 n_3  &\geq \ell_3 -1. \label{eq:cycle4}
\end{align}
From \cref{eq:cycle1}, \cref{eq:cycle3}  and \cref{eq:cycle4}, we obtain
 \begin{align}
  2(n_1 + n_3)& \geq g + 2 \lceil{(\ell_1-1)/2}\rceil - \lfloor{\ell_1/2}\rfloor -1 \\
  &= g + \lceil{(\ell_1-1)/2}\rceil - 1\\
  &\geq 3g/2-1.
 \end{align}
 If $n > n_1+n_3$ (that is, there is some GREEN edge outside $C \cup P^*$),
 then we obtain $n\geq n_1+n_3+1\geq 3g/4+1/2$, contradicting our assumption $n\leq 3g/4$.
 So, we may assume that all GREEN edges in the graph are contained in $C \cup P^*$. We think of $C\cup P^*$ as a set of three edge disjoint paths, $P_1$, $P_2$ and $P_3$, connecting $r_1$ to $r_2$, where $P_1 \cup P_2 = C$ and $P_3=P^*$. Let $V^*$ be the set of vertices in $P_1 \cup P_2 \cup P^*$. We will show that the graph induced by $V^*$ can be safely oriented. Then, we will orient all other (necessarily RED) edges of $G$ towards a vertex not in $V^*$ to obtain a safe orientation of the \emph{entire} graph $H$, and conclude that the lemma holds with $V'=V(H)$. 
 
 First, we show that we may assume that each of the three cycles $C=P_1 \cup P_2$, $P_1 \cup P_3$ and $P_2 \cup P_3$ is chordless. Since $C$ is GREEN-dominated, it has at most $2n+1$ edges. If it had a chord, we would get a cycle with at most $n\leq 3g/4 < g$ edges, a contradiction. Thus, the cycle $C = P_1 \cup P_2$ has no chord. Next, using a similar argument we show that the other two cycles are chordless. We first observe that both $P_1$ and $P_2$ have GREEN edges. If $P_1$ has no GREEN edges, then it can have at most two edges, and the at least $g/2$ GREEN edges of $C$ all lie in $P_2$.  Also, $P_3$ has at least $g-2$ edges. Then, the number of GREEN 
 edges in $P_2$ is at least $\max\{|P_2|,g\}/2$  (because $C$ is GREEN-dominated), and similarly the number of GREEN edges in $P_3$ is at least $\lceil{(|P_3|-1)}/2\rceil \geq g/2-1$. Thus,
 \begin{equation}
 3/4g \geq n \geq  \max\{|P_2|,g\}/2 + \lceil{(|P_3|-1)/2}\rceil \geq g-1;
 \end{equation}
 that is, $g=4$, $|P_2|\leq 4$ and $|P_3| \leq 3$. Thus,
 $P_2 \cup P_3$ is a cycle with at most $7$ vertices and it cannot have a chord because $g=4$.  Thus, we may assume that $P_1$ has has at least one GREEN edge, that is, $P_2 \cup P_3$ has at most $n-1$ GREEN edges. Let $k_2$ be the number of GREEN edges in $P_2$ and $k_3$ the number of GREEN edges in $P_3$. Since every RED edge in $P_3$, except perhaps one is followed by a GREEN edges, the number of edges in $P_3$ is at most $2k_2+2$. Then, the total number of edges
 in $P_2 \cup P_3 \leq (2k_2+2)+(2k_3+1)$ (the second term comes from \cref{eq:cycle4}), that is, at most $2n+1$ edges in all. If $P_2 \cup P_3$ has a chord, then we have a cycle of length at most $n$, which, as we saw earlier, is not possible. Similarly, $P_2 \cup P_3$ has not chord. 
 
 So the graph induced by $V^*$ consists of three disjoint paths, with no chords across them.  If any path has two consecutive RED edges, then we may orient them towards each other and be left with a graph consisting of a cycle with two dangling paths, which can be oriented safely. Similarly, if some two paths start with RED edges or end with RED edges, then these edges can be oriented towards each other, and the remaining edges (which form a tree) can be oriented safely.  We are left with the case where on all paths a RED edge is followed by a GREEN edge, and at both ends ($r_1$ and $r_2$), two of the paths start with GREEN edges. We will show that this is impossible. For otherwise, there is path (say, $P_3$), which has GREEN edges at both ends, so $|P_3|$ has at least $(|P_3|+1)/2$ GREEN edges. For the remaining paths, either some path has both ends GREEN, or both paths have one end GREEN. In either case, they together have at least $(|P_1| + |P_2|)/2$ GREEN edges. Note that $2(|P_1|+|P_2|+|P_3|) \geq 3g$, because $H$ has girth at least $g$. Thus, the total number of GREEN edges is
 $n \geq (|P_1| + |P_2| + |P_3| + 1)/2 \geq (3g/2 +1)/2 > 3g/4,$
 contradicting our assumption that $n\leq 3g/4$. 
 \end{proof}
\section{Quantum adaptive scheme}
In this section, we establish \cref{thm:result2}.
Our quantum scheme is based on a graph and is similar in some respects to the classical scheme described above. The main difference is in the second probe, which now computes the XOR of two bits of memory.
\begin{definition}[Quantum $(G,K)$-scheme]
Let $G$ be a directed graph with $N$ vertices and $M$ edges; let $K$ be a positive integer.
We refer to the following as a quantum $(G,K)$-scheme. The storage consists of three bit arrays, $A$, $B_0$ and $B_1$. To answer a membership query, the quantum decision tree  first probes array $A$ (this probe is classical) and 
then computes the XOR of two bits in either $B_0$ or $B_1$, using just one quantum probe.
\begin{description}
\item[Edge array:] An array $A: E(G) \rightarrow \{0,1\}$, indexed by edges of $G$. 
\item[Vertex arrays:] Two  two-dimensional arrays $B_0,B_1: V \times [K] \rightarrow \{0,1\}$, indexed by elements of the form $(v,i)$. 
\item[Elements:] As before, we identify our universe of elements $[m]$ with a subset of $E(G) \times [K]$; thus, each element $x \in [m]$ will be referred to as $(e,i)$. 
\item[Query:] Let the query be ``Is $x$ in $S$?'', where $x=(e,i)$; suppose $e=\{x_0,x_1\}$. To process this query for we read $A[e]$ (first probe); then, based on
the value of $A[e]$, we return either $B_0[(v_0,i)] + B_0[(v_1,i)] \pmod{2}$ or $B_1[(v_0,i)] + B_1[(v_1,i)] \pmod{2}$. In other words, the first probe directs us to either array $B_0$ or $B_1$; we then return the XOR of the bits in the $i$-th location in the rows corresponding to the two vertices of $e$. 
\item[Space:] We will ensure that $MK \geq m$,  to accommodate all elements of the universe. The space used by this scheme is then $2NK + M$ bits. By choosing the graph $G$ and the parameter $K$ appropriately we will show that our schemes uses small space.
\end{description}
\end{definition}
The main idea is the following. To store the set $S$ in the data structure, we partition the edges of $G$ using the $0$-$1$ assignment to the array $A$. Let $G_0$ be the graph induced by the edges that are assigned $0$ in $A$, and let $G_1$ be the graph induced by the edges assigned $1$. Now, the bits of the arrays $B_0$ and $B_1$ must be assigned in such a way that certain XORs of two bits evaluate to $1$ and others evaluate to $0$. This leads to a system of linear equations in the bits of the arrays $B_0$ and $B_1$. To ensure that this system has a solution, we ensure that if $A[e]=0$, then $e$ is not in any cycle in $G_0$, and similarly, if $A[e]=1$, then $e$ is not in any cycle in $G_1$. It is then easy to see that the required assignment to the arrays $B_0$ and $B_1$ exists. To ensure that the edges of $G$ can be partitioned in $G_0$ and $G_1$ to satisfy the requirements imposed by the set $S$, we will start with the graph $G$ that is dense but locally sparse in the following sense, and use a theorem of Nash-Williams.
\begin{definition}[Locally sparse graph]
A graph $G$ is $(k,\alpha)$-locally sparse if for every subsets $V' \subseteq V$ with $4 \leq |V'| \leq k$ vertices, the induced subgraph on $V'$ has at most $\alpha |V'|$ edges.
\end{definition}
\begin{lemma} \label{lm:main_quantum_adaptive}
If $G$ has $N$ vertices, $M$ edges and is $(4n,5/4)$-locally sparse, then 
\[s_{QA}(m,n,t=2)\leq M + 2N\lceil\frac{m}{M}\rceil.\]
\end{lemma}
Before we present the proof of this lemma, let us see how this leads to \cref{thm:result2}. We will need a family of 
dense locally sparse graphs, whose existence we establish in \cref{appendix:probabilistic} using a routine probabilistic argument.
\begin{lemma} \label{lm:locally_sparse_construction}
For all large $N$ there is a $(4N^{1/6},5/4)$-locally sparse graph with $N$ vertices and $\Omega(N^{7/6})$ edges.
\end{lemma}
We set $N=m^{3/4}$, and plugging in the graph promised by  \cref{lm:locally_sparse_construction} in \cref{lm:main_quantum_adaptive}, obtain $s_{QA}(m, m^{1/8},2) = O(m^{7/8})$, as claimed in \cref{thm:result2}. It remains to establish \cref{lm:main_quantum_adaptive}.
\begin{proof}[Proof of \cref{lm:main_quantum_adaptive}]
Fix $G$ with the given parameters. We now describe how the three arrays in our quantum scheme are assigned values. 
Recall that we view elements of $[m]$ as pairs $(e,i)$. 
Edges of $G$ for which there is an element of the form $(e,i) \in S$ will be called GREEN; the other edges of $G$ will be called RED. Say, there are $\ell \leq n$ GREEN edges. We will construct a sets of vertices $D_0, D_1, D_2, \ldots$ by 
adding one vertex at a time. Let $D_0 \subseteq V(G)$ be the union of the GREEN edges; thus $D_0$ has at most $2\ell$ vertices. To obtain $D_{i+1}$ from $D_i$, add to $D_i$ a new vertex that has at least two edges leading into $D_i$; if no such vertex exists, stop. We claim that this process stops before $2n$ vertices are added, for otherwise,  the graph induced by $D_{2n}$, a set of size at most $2n+2\ell \leq 4n$ vertices, has at least $4n+\ell$ edges. Since $G$ is $(4n,5/4)$-locally sparse, we have $(5/4)(2n+2\ell) > 4n+\ell$, implying $\ell > n$, a contradiction. Let $D$ be the set of vertices when the above process stops. 
\begin{claim} \label{cl:nash-williams}
The subgraph induced by $D$ can be split into two disjoint forests.
\end{claim}
We will justify this claim below (using Nash-Williams theorem). Let us assume it and 
complete the proof. Let the two forests guaranteed by this claim be $F_1$ and $F_2$. We set $A[e]=0$ for all edges $e \in F_1$ and all edges that connect $D$ to $V\setminus D$. Let $G_0$ be the subgraph of $G$ with vertex set $V(G)$ that consist of edges $e$ such that $A[e]=0$. Let $G_1$ be the subgraph with vertex set $V(G)$ and all edges not included in $G_0$. Note that the connected components of $G_1$ are either in the forest $F_2$ or consist of RED edges with both end points in $V\setminus D$. Now, we are ready to describe the assignment to arrays $B_0$ and $B_1$. As stated above the constraints imposed by the GREEN and RED edges give a system of equational constraints; since $G_0$ has no cycle, it is easy to see that these constraints can all be satisfied by assigning $B_0$ appropriately. In $G_1$ again, the edges corresponding to $F_2$ do not induce a cycle, so the constraints imposed by them can be satisfied by 
assigning appropriate values to the rows of $B_2$ corresponding to vertices in $D$. The remaining edges share no vertex with the edges of $F_2$, and consist only of RED edges. So we assign zeroes to all rows of $B_2$ corresponding to vertices in $V(G) \setminus D$. 

It remains to verify \cref{cl:nash-williams}. Since $|D| \leq 4n$, every subset $D'$ of $D$ (with $|D'|\geq 4$) induces at most $(5/4)|D'|$ edges; since $|D| \geq 4$, we have $(5/4)|D'| \leq 2(|D'| -1)$. Note that the number of edges in any graph with at most $1\leq \ell \leq 4$ vertices is at most $2\ell-2$. So we may invoke \cref{thm:nash} below and 
justify the claim. \end{proof}
\begin{theorem}[Nash-Williams (see Theorem 3.5.4 of \cite{diestel2000}).] \label{thm:nash}
Let $H=(V,E)$ be an undirected graph such that for each non-empty subset $X\subseteq V$, the number of edges with both end points in $X$ is at most $2(|X|-1)$. Then $E$ can be partitioned as $E=E_1 \cup E_2$ such that $(V,E_1)$ and $(V,E_2)$ are both forests.
\end{theorem}

\section{Quantum non-adaptive schemes for $n=2$ and $t=2,3$}

In this section, we show that the lower bound in \ref{eq:garg:radhakrishnan} is tight for two cases: $s_{QN}(m,n=2,t=2) =O(\sqrt{m})$ and $s_{QN}(m,n=2,t=3) = O(m^{1/3})$; the schemes we give are non-adaptive and only use the fact that the XOR of two bits can be computed using one quantum query. The proofs are algebraic.

\subsection{Case $t=2$}

We identify $[m]$ with $A \times B$, where each of the sets has about $\sqrt{m}$ elements; $A$ and $B$ are disjoint. We will have two array indexed by $A$ (we call them $X_1$ and $X_2$) and two arrays indexed by $B$ (we call them $Y_1$ and $Y_2$). 
\begin{description}
\item[Query:] On receiving the element $x=(x_1,x_2) \in A \times B$, the algorithm returns 
\[ (X_1[x_1]+Y_1[x_2])(X_2[x_1]+Y_2[x_2]) \pmod{2},\]
which is a polynomial of degree two. Note that both $X_1[a]+Y_1[b]$ and $X_2[a]+Y_2[b]$ can be computed in parallel with one quantum query each. Thus, the scheme requires only two non-adaptive queries. 

\item[Storage:] Given a pair of elements $\{\alpha_1,\alpha_2\} \subseteq [m]$, we need to show how values will be assigned to the four arrays: $X_1,X_2,Y_1,Y_2$. It will be easier to describe and analyse our storage algebraically. We view $X_1,X_2$ as functions from $A$ to $\{0,1\}$ and $Y_1,Y_2$ as functions from $B$ to $\{0,1\}$. For $a \in A$, let $\delta_a: A \rightarrow \{0,1\}$  be defined by $\delta_a(z)=1$ iff $z=a$; similarly for $b \in B$, let $\delta_b: B \rightarrow \{0,1\}$ be defined by $\delta_b(z)=1$ iff $z=b$. We have three cases based on the number of components $\ell \in \{0,1,2\}$ where $\alpha_1$ and $\alpha_2$ agree. 
\end{description}

\begin{description}
\item[$\ell=2$:] We have only one element $(a,b)$. We set $X_1\equiv \delta_a$, $Y_1=0$, $X_2\equiv 0$ and $Y_2\equiv \delta_b$. The query polynomial reduces to the monomial $\delta_a(x_1) \delta_b(x_2)$, which is what we want.
\item[$\ell=1$:] Say the set is $\{(a,b), (a',b)\}$. We set $X_1\equiv \delta_a + \delta_{a'}$, $Y_1\equiv 0$, $X_2 \equiv 0$ and $Y_2\equiv\delta_b$. The query polynomial reduces to $(\delta_a(x_1)+\delta_{a'}(x_2))\delta_b(z_2) = \delta_a(z_1)\delta_b(z_2) + \delta_{a'}(z_1)\delta_{b}(z_2)$, which is what we want.
\item[$\ell=0$:] Say the set is $\{(a,b), (a',b')\}$. We set $X_1\equiv \delta_a$, $Y_1\equiv \delta_{b'}$, $X_2 \equiv \delta_{a'}$ and $Y_2\equiv \delta_b$. The query polynomial evaluates $(\delta_a(z_1)+\delta_{b'}(z_2))(\delta_{a'}(z_1)+\delta_b(z_2))= \delta_a(z_1)\delta_b(z_2)+\delta_{a'}(z_1)\delta_{b'}(z_2) \pmod{2}$, which is what we want.
\end{description}

\subsection{Case $t=3$}
Let us identify $[m]$ with $A\times B \times C$, where each of the sets has roughly $m^{1/3}$ elements; we will assume that $A$, $B$ and $C$ are disjoint. 
We have six arrays, two indexed by $A$ (we call them $X_1$ and $X_2$), two indexed by $B$ (we call them $Y_1$ and $Y_3$) and two indexed by $C$ (we call them $Z_2$ and $Z_3$); the subscripts indicate which query probes the corresponding array. 
\begin{description}
\item[Query:] On receiving the element $e=(x,y,z)$, the algorithm returns 
\[ (X_1[x]+Y_1[y])(X_2[x]+Z_2[z])(Y_3[y]+Z_3[z]) \pmod{2},\]
which is a polynomial of degree $3$.

\item[Storage:] Given a pair of elements $\{\alpha_1,\alpha_2\}$, we need to show how values will be assigned to the six arrays. Let $\alpha=(a,b,c)$ and $\beta=(a',b',c')$. We define functions of the form $\delta_a:A\rightarrow\{0,1\}$, $\delta_b:B\rightarrow \{0,1\}$ and $\delta_c:C\rightarrow \{0,1\}$ as before.
Also $0$ and $1$ when denoting functions correspond to the all $0$'s and the all $1$'s functions.
%
We have four cases, depending on the number of places $\ell \in \{0,1,2,3\}$ where $\alpha_1$ and $\alpha_2$ agree.
\end{description}
\begin{description}
\item[$\ell=3$:] The set has only one element $(a,b,c)$, say. The arrays are as follows.
$X_1 \equiv \delta_a$; $Z_2\equiv \delta_c$; $Y_3\equiv \delta_b$, the other three arrays are $0$. So, the query function becomes $(\delta_a(x)+0)(\delta_c(z)+0)(\delta_b(y)+0) = \delta_a(x)\delta_b(y)\delta_c(z)$, which yields 1 iff $(x,y,z)=(a,b,c)$.

\item[$\ell=2$:] Say $\alpha_1=(a,b,c)$ and $\alpha_2=(a,b,c')$. We set 
$X_1\equiv \delta_a$, $Y_1\equiv 0$, $X_2 \equiv 0$, $Z_2\equiv \delta_c + \delta_{c'}$, 
$Y_3\equiv \delta_b$ and $Z_3\equiv 0$. Then, the query function becomes
$(\delta_a(x)+0)(\delta_{c}(z)+\delta_{c'
}(z))(\delta_b(y)+0)$, which reduces to $\delta_a(x)\delta_b(y)\delta_c(z)+
\delta_a(x)\delta_b(y)\delta_{c'}(z)$, that is, the function that evaluates to $1$ precisely when the input is $(a,b,c)$ or $(a,b,c')$. The other cases are symmetric.

\item[$\ell=1$:] Say $\alpha_1=(a,b,c)$ and $\alpha_2=(a,b',c')$.
We set $X_1 \equiv 1+\delta_a$, $Y_1 \equiv \delta_b+\delta_{b'}$, $X_2\equiv 1+\delta_a$, $Z_2 = \delta_c+\delta_{c'}$, $Y_3=\delta_b$ $Z_3=\delta_{c'}$. 
Our query polynomial then evaluates to 
\begin{equation}
 (1+\delta_a+\delta_b+\delta_{b'}) (1+\delta_a+\delta_c+\delta_{c'}) (\delta_b+\delta_{c'}), \label{eq:query1}
 \end{equation}
where, to simplify notation, we just write $\delta_a$ instead of $\delta_a(x)$, etc.
Applying the rule $gh =(g+h+1)h$ twice, we obtain
$
(\delta_c+\delta_{b'})(1+\delta_a + \delta_c+\delta_{c'})(\delta_b+ \delta_{c'})$.
Then, combining the first and last factors, we obtain,
$
(\delta_b\delta_c+\delta_{b'}\delta_{c'})(1+\delta_a + \delta_c+\delta_{c'}).
$
Expanding this mod 2, we obtain
$(\delta_b\delta_c+\delta_{b'}\delta_{c'}) \delta_a$, which
yields $1$ iff $x \in \{(a,b,c), (a,b',c')\}$, as required.


\item[$\ell=0$ (the two elements differ on all coordinates):]  
We set $X_1\equiv \delta_a$,  $X_2\equiv \delta_{a'}$, $Y_1\equiv \delta_{b'}$, $Y_2\equiv \delta_{b}$, $Z_1\equiv \delta_{c'}$, $Z_2\equiv \delta_{c}$. The query expression evaluates to
\[ (\delta_a+ \delta_{b'})(\delta_{a'}+\delta_{c})(\delta_{b}+\delta_{c'}).\]
Focus on the middle factor. If we pick $\delta_{a'}$ from that factor, then we are forced to pick $\delta_{b'}$ from the previous, which forces us to pick $\delta_{c'}$ from the last (to avoid getting 0); if we pick $\delta_{c}$ from the middle factor, then we are forced to pick $\delta_{b}$ from the last and then $\delta_{a}$ from the first. All other terms are 0. 
The final expression with two terms is $\delta_a \delta_b \delta_c + \delta_{a'}\delta_{b'}\delta_{c'}$, as required.
\end{description}

\bibliography{references}

\appendix
\section{Explicit construction of graphs}
\label{sec:appendix_explicit}
We say that a graph on $L$ vertices is explicitly, if the adjacency matrix of $L$ can be constructed in polynomial time in $L$.
\begin{itemize}
\item We will use the construction due to Wenger~\cite{WENGER1991113} to exhibit explicit graphs with girths $8$. Wenger constructs a graph $H_k(p)$ with $2p^k$ vertices and $2p^{k+1}$ edges, and shows that
if $p$ is prime, then $H_3(p)$ has girth at least $8$.  In the bipartite graph $H_k(p)$ the vertices are represented as $k$-tuples of numbers $\{0,\ldots,p\}$ and two vertices are connected based on a simple arithmetic check involving addition and multiplication modulo $p$. In our application, given a number $L$, we set $p$ to be the smallest prime that is at least $L^{1/(k+1)}$. Then, $H_k(p)$ has at most $2^{k+2} L$
vertices and at least $2L^{1+1/k}$ edges. Thus, we obtain graphs with the following parameters: $g=8$, 
$c(g=8)= 2^{5}$, $\tau(g=8) = 1/3$. Our application for $g=8$ (see \cref{fig:comparison}) uses these parameters.

\item For girth 12, we use a construction due to Benson~\cite{benson_1966}. Theorem 2 of the paper presents 
an algebraic construction where the graph is obtained by considering point-line incidences for points and lines of a quardic surface in the projective $6$-space $P(6,q)$. The degree of each vertex of this graph is $q+1$. On page 1093, the number of vertices in this graph is computed to be
$(q+1)(1+q^2+q^4)$. So, to get the graph suitable for our applications, we take $q$ to be the smallest prime such that 
$(q+1)(1+q^2+q^4)\geq L$ and use this construction. Then, it is easy to see that the number of vertices in this graph is $\mathcal{O}(L)$ and the number of edges is at least $L^{1+1/5}$.


\item Lazebnik, Ustimenko and Woldar~\cite{girthpaper} exhibit dense graphs for various values of girth. Their Corollary 3.3 shows
graphs with girth at least $2s+2$, with  $v \leq 2 q^{(3s-3)/2}$
vertices if $s$ is odd and at most $2q^{(3s-2)/2}$ vertices if $s$ is even.  The graph has $\frac{1}{2}vq$ edges. To construct the graphs for our application, fix $L$ and let $q$ be the smallest prime larger than $L^{2/(3s-3)}$ or $L^{2/(3s-2)}$
(depending on whether $s$ is odd or even) and consider the graph obtained from the above construction. If the graph has fewer than $L$ vertices, then we put together disjoint copies of it, to obtain one with number of vertices between $L$ and $2L$. It can be verified that this graph has $O(L)$ vertices and $\Omega(L^{1+2/(3s-3)})$ or $\Omega(L^{1+2/(3s-2)})$ edges (depending on whether $s$ is odd or even). In our application (see \cref{fig:comparison}),  we use graphs with girth $4r$ and $4r-2$. Setting $2s+2=4r$, i.e., $s=2r-1$ (an odd number), we obtain a graph with  $\Omega(L^{1+1/(3r-3))})$ edges; similarly setting $2s+2=4r-2$, i.e., $s=2r-2$ (an even number), we obtain a graph with $\Omega(L^{1+1/(3r-4)})$ edges.
\end{itemize}

\section{Examples that show {\cref{lm:safe}} is tight}

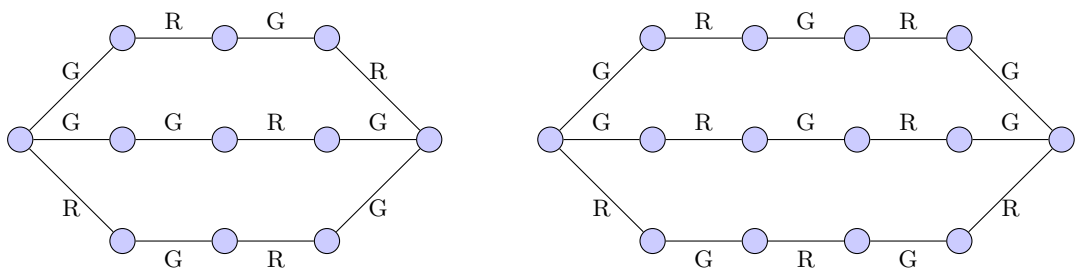
\begin{figure}[t]
    \begin{tikzpicture}
\matrix[matrix of math nodes, anchor= west,
      nodes={circle, fill = blue!20, draw, minimum size = 0.2 cm},
        column sep = {1cm},
        row sep={1 cm}]
{
 &          & |(a)|    & |(b)|    & |(c)|  & |(d)|    \\
&|(s)|  & |(e)|    & |(f)|    & |(g)|  & |(k)| & |(t)|  \\
&         & |(h)|    & |(i)|    & |(j)|   & |(l)| & \\
};
 \foreach \from/\to/\weight/\where 
         in {s/a/G/above, s/e/G/above, s/h/R/below, a/b/R/above,  b/c/G/above, c/d/R/above, d/t/G/above, e/f/R/above, f/g/G/above,  g/k/R/above, k/t/G/above, h/i/G/below, i/j/R/below, j/l/G/below,l/t/R/below}
    \draw (\from) to [->]  node[midway,\where]  {\small{\weight}}  (\to);
\matrix[matrix of math nodes, anchor= east,
      nodes={circle, fill = blue!20, draw, minimum size = 0.2 cm},
        column sep = {1 cm},
        row sep={1 cm}]
{
           & |(a)|    & |(b)|   & |(d)|    \\
    |(s)|  & |(e)|    & |(f)|   & |(k)| & |(t)|  \\
         & |(h)|    & |(i)|     & |(l)| & \\
};
 \foreach \from/\to/\weight/\where 
         in {s/a/G/above, s/e/G/above, s/h/R/below, a/b/R/above,  b/d/G/above, d/t/R/above, e/f/G/above, f/k/R/above,   k/t/G/above, h/i/G/below,  i/l/R/below,l/t/G/below}
    \draw (\from) to [->]  node[midway,\where]  {\small{\weight}}  (\to);
\end{tikzpicture}
    \caption{Examples of graphs with girth $g=10$ and $g=8$ that cannot be safely oriented}
    \label{fig:tight}
\end{figure}

The bound shown \cref{lm:safe} is tight in the following sense: for each positive even integer $g$, there exists a bipartite graph with girth $g$ and  $\lfloor{4g/3}\rfloor+1$ GREEN edges that cannot be safely oriented. For example, the graph consisting of three edge-disjoint $s$-$t$ paths, 
each of length $2k$, has girth $g=4k$; but we can designate a set of $n=3k+1$ edges GREEN for which the graph has no safe orientation. For this graph $n=3k+1$ and $\lfloor{3g/4}\rfloor = 3k$. 
A similar example, with three edge-disjoint paths of length $2k+1$,  shows that the above lemma is tight for $g=4k+2$. 
\cref{fig:tight} shows these examples for $k=2$.

\section{Proof of \cref{lm:locally_sparse_construction}}
\label{appendix:probabilistic}
Consider the random graph on $N$ vertices where each edge 
is picked independently with probability $p=(1/50)n^{-5/6}$. The probability that $G$ is not ($4N^{1/6}$, $5/4$)-locally sparse is at most (we use the union bound over the choice of subsets of size $\ell \leq  4N^{1/6}$, and for each set over all choice of
$1.24\ell$ edges
for simplicity we ignore floors and ceilings):
\begin{align*}
    \sum_{\ell=5}^{4N^{1/6}} {N \choose \ell}{\ell^2 \choose {5/4 \ell}} p^{5/4 \ell} &\leq
    \sum_{\ell=5}^{4N^{1/6}} \left(\frac{eN}{\ell}\right)^\ell
    \left(\frac{e\ell^2}{5/4\ell}\right)^{{5/4}\ell} p^{5/4 \ell}
    \\
    &\leq\sum_{\ell=5}^{4N^{1/6}} 8^\ell \left(N\ell^{1/4}\right)^\ell p^{(5/4)\ell} \\
    &\leq \sum_{\ell=5}^{4N^{1/6}} \left(\frac{1}{8}\right)^\ell \left(N\ell^{1/4}\right)^{\ell}  N^{-(5/6)(5/4)\ell}.
\end{align*}
By considering terms corresponding to (say) $\ell < N^{1/12}$ and $\ell \geq N^{1/12}$ separately, we see that the last sum is $o(1)$. Thus, with high probability no set of size up to $4 N^{1/6}$ violates the local sparsity condition. Furthermore, with high probability the number of edges in the graph is at least $pN^2/2 = \Omega(N^{7/6})$. Thus, there exists an $(4N^{1/6},5/4)$-sparse graph with
at $\Omega(N^{7/6})$ edges.

\section{Non-adaptive quantum bounds} 
\label{appendix:non-adaptive}
We give an upper bound on $s_Q(m,n = 2k+1,t=2)$ for the non-adaptive classical scheme, where $k \in \mathbb{N}$. The arrangement of the element and bits is similar to the classical adaptive scheme we described in \cref{sec:classical_adaptive}. The first probe is on the corresponding edge array and the second is an equality probe on the rows corresponding to the vertices of the edge. We AND the two probes to answer membership queries. We obtain
\begin{equation}
    s_Q(m,n = 2k+1,t=2) = \begin{cases}
        \mathcal{O}(v^{1+\frac{4}{3n-9}}) & \text{if }  4 |(n + 1) ;\\
        \mathcal{O}(v^{1+\frac{4}{3n-7}})  & \text{if }  4 \nmid (n+1).
    \end{cases}
\end{equation}

\begin{definition}[Non-adaptive Quantum $(G,K)$-scheme]
Let $G$ be an un-directed graph with $N$ vertices and $M$ edges; let $K$ be a positive integer.
We refer to the following as a (non adaptive) quantum $(G,K)$-scheme. The storage consists of two bit arrays, $A$ and $B$. To answer a membership query the decision tree will make the first probe to array $A$ and the second probe to array $B$.
\begin{description}
\item[Edge array:] An array $A: E(G) \rightarrow \{0,1\}$, indexed by edges of $G$. 
\item[Vertex array:] A two dimensional array $B: V \times [K] \rightarrow \{0,1\}$. 
\item[Elements:] We identify our universe of elements $[m]$ with
a subset of $E(G) \times [K]$ (so we must ensure that the graph has
at least $m/K$ edges); thus, each element $x \in [m]$ will be referred to as $(e,i)$. 

\item[Query:] We represent an edge of $G$ as an ordered pair of the form $e=(v_0,v_1)$. To process the query for the element $x=(e,i)$, we return the value $A[e] \cdot (B[v_0,i]\oplus B[v_1,i])$. 

\item[Space:] We will ensure that $MK \geq m$. The space used by this scheme is then $NK + M$ bits ($NK$ for the $N$ vertex array and $K$ for the edge array). By choosing the graph $G$ and the parameter $K$ appropriately we will show that our schemes use small space.
\end{description}
\end{definition}
 
As in the classical adaptive setting, an edges $e$ is coloured GREEN if $(e,i) \in S$ for some $i$. Values can be assigned consistently to the arrays if there is no cycle in the graph consisting entirely of GREEN edges. This idea is formalized in the lemma below.

\begin{lemma}
Let  $G$ be a graph with $N$ vertices $M$ edges and girth $g$ such that $n < g$. Then, $s_{A}(m,n,2) \leq M + N\lceil m/M \rceil$.
\end{lemma}

\end{document}